\documentstyle[psfig]{mn}

\newif\ifAMStwofonts

\def\simlt{\lower.5ex\hbox{$\; \buildrel < \over \sim \;$}}
\def\simgt{\lower.5ex\hbox{$\; \buildrel > \over \sim \;$}}

\ifoldfss
  \ifCUPmtlplainloaded \else
    \NewTextAlphabet{textbfit} {cmbxti10} {}
    \NewTextAlphabet{textbfss} {cmssbx10} {}
    \NewMathAlphabet{mathbfit} {cmbxti10} {} 
    \NewMathAlphabet{mathbfss} {cmssbx10} {} 
  \fi
  \ifAMStwofonts
    \ifCUPmtlplainloaded \else
      \NewSymbolFont{upmath} {eurm10}
      \NewSymbolFont{AMSa} {msam10}
      \NewMathSymbol{\upi}     {0}{upmath}{19}
      \NewMathSymbol{\umu}     {0}{upmath}{16}
      \NewMathSymbol{\upartial}{0}{upmath}{40}
      \NewMathSymbol{\leqslant}{3}{AMSa}{36}
      \NewMathSymbol{\geqslant}{3}{AMSa}{3E}

       \let\ge=\geqslant
    \fi
  \fi
\fi 

\ifnfssone
  \newmathalphabet{\mathit}
  \addtoversion{normal}{\mathit}{cmr}{m}{it}
  \addtoversion{bold}{\mathit}{cmr}{bx}{it}
  \newmathalphabet{\mathbfit} 
  \addtoversion{normal}{\mathbfit}{cmr}{bx}{it}
  \addtoversion{bold}{\mathbfit}{cmr}{bx}{it}
  \newmathalphabet{\mathbfss} 
  \addtoversion{normal}{\mathbfss}{cmss}{bx}{n}
  \addtoversion{bold}{\mathbfss}{cmss}{bx}{n}
  \ifAMStwofonts
    \ifCUPmtlplainloaded \else
      \UseAMStwoboldmath
      \makeatletter
      \new@mathgroup\upmath@group
      \define@mathgroup\mv@normal\upmath@group{eur}{m}{n}
      \define@mathgroup\mv@bold\upmath@group{eur}{b}{n}
      \edef\UPM{\hexnumber\upmath@group}
      \new@mathgroup\amsa@group
      \define@mathgroup\mv@normal\amsa@group{msa}{m}{n}
      \define@mathgroup\mv@bold\amsa@group{msa}{m}{n}
      \edef\AMSa{\hexnumber\amsa@group}
      \makeatother
      \mathchardef\upi="0\UPM19
      \mathchardef\umu="0\UPM16
      \mathchardef\upartial="0\UPM40
      \mathchardef\leqslant="3\AMSa36
      \mathchardef\geqslant="3\AMSa3E

       \let\ge=\geqslant
    \fi
  \fi
\fi 

\ifnfsstwo
  \DeclareMathAlphabet{\mathbfit}{OT1}{cmr}{bx}{it}
  \SetMathAlphabet\mathbfit{bold}{OT1}{cmr}{bx}{it}
  \DeclareMathAlphabet{\mathbfss}{OT1}{cmss}{bx}{n}
  \SetMathAlphabet\mathbfss{bold}{OT1}{cmss}{bx}{n}
  \ifAMStwofonts
    \ifCUPmtlplainloaded \else
      \DeclareSymbolFont{UPM}{U}{eur}{m}{n}
      \SetSymbolFont{UPM}{bold}{U}{eur}{b}{n}
      \DeclareSymbolFont{AMSa}{U}{msa}{m}{n}
      \DeclareMathSymbol{\upi}{0}{UPM}{"19}
      \DeclareMathSymbol{\umu}{0}{UPM}{"16}
      \DeclareMathSymbol{\upartial}{0}{UPM}{"40}
      \DeclareMathSymbol{\leqslant}{3}{AMSa}{"36}
      \DeclareMathSymbol{\geqslant}{3}{AMSa}{"3E}

       \let\ge=\geqslant
    \fi
  \fi
\fi 

\ifCUPmtlplainloaded \else
  \ifAMStwofonts \else 
    \def\upi{\pi}
    \def\umu{\mu}
    \def\upartial{\partial}
  \fi
\fi

\title[Annihilation of Cold Dark Matter]
    { The Clumpiness of  Cold Dark Matter: Implications for the Annihilation Signal}
\author[Taylor \& Silk]
{James E. Taylor and Joseph Silk\\
Denys Wilkinson Building, 1 Keble Road, Oxford OX1 3RH, United
Kingdom}
\date{Draft version \today}
\pubyear{2002}

\begin{document}

\maketitle

\begin{abstract}
We examine the expected signal from annihilation events in realistic 
cold dark matter halos. If the WIMP is a neutralino, with an annihilation 
cross-section predicted in minimal SUSY models for the lightest stable 
relic particle, the central cusps and dense substructure seen in simulated 
halos may produce a substantial flux of energetic gamma rays. We derive 
expressions for the relative flux from such events in simple halos with 
various  density profiles, and use these to calculate the relative flux 
produced within a large volume 
as a function of redshift. This flux peaks when the first halos collapse, 
but then declines as small halos merge into larger systems of lower density. 
Simulations show that   halos contain a substantial amount of dense 
substructure, left over from the incomplete disruption of smaller halos 
as they merge together. We calculate the contribution to the flux due to 
this substructure, and show that it can increase the annihilation signal
substantially. Overall, the present-day flux from annihilation 
events may be an order of magnitude larger than predicted by 
previous calculations. We discuss the implications of these results 
for current and future gamma-ray experiments. 
\end{abstract}



\section{Introduction}

Dark matter is omnipresent in the universe.
Most of it is non-baryonic, and a favoured candidate is a
weakly interacting massive particle (WIMP),
often generically taken to be the lightest stable relic particle
surviving from when the universe was supersymmetric. The freeze-out of 
such a neutralino of mass $m_\chi$ occurs at 
$kT\sim m_\chi/20$, 
and the annihilation cross-section determines the  
current value of the CDM density $\Omega_{\rm cdm}h^2.$ For minimal SUSY, for 
example, one can derive the relation between annihilation cross-section 
and particle mass,
and compute the various annihilation products, including continuum
gamma rays from
$\pi^0$ decays and line gamma rays from rare quark decays
(c.f.\ Bergstr\"om 2000 for a recent review). 
While cosmological observations specify 
$\Omega_{\rm cdm}h^2\approx 0.1,$
scanning over minimal SUSY parameter space 
results in an uncertainty in the gamma-ray emissivity of several orders of 
magnitude.

Our galactic halo is a logical place to look for evidence of annihilations.
Unfortunately, for a uniform dark halo with a realistic density profile, 
even the most optimistic models fall short of the observed diffuse 
high-galactic-latitude gamma-ray flux, as measured by EGRET, 
by an order of magnitude or more (Ullio et al.\ 2002). 
In fact, high-resolution numerical simulations show that the dark halo has 
considerable substructure (Klypin et al.\ 1999; Moore et al.\ 1999). 
This substructure may boost the annihilation flux substantially. If this 
is indeed the case, then the isotropic diffuse 
background flux from the many small halos that merged in the past into 
our halo and others may also become significant. There is considerable 
uncertainty among cosmological halo simulators, however, about the quantitative 
role of substructure in our dark halo, and of the concentration of the 
substructure and of the dark matter itself towards the centre of the galaxy.

No reliable estimates have been given up till now of the properties of 
halo substructure on very small scales, and in particular of their 
dependence on halo mass and their evolution with cosmological epoch. 
We have developed a semi-analytical model of halo formation which is capable of 
following the key physics of tidal disruption of substructure during merging.
In principle this approach has arbitrarily high resolution.
Hence we are able to provide robust calculations of the annihilation flux
generated within our own halo, but also especially of the component
generated during the evolution of structure at early times,
and visible as an isotropic gamma-ray background at the present day.

In this paper, we calculate the flux produced by annihilations in 
simple CDM halos, relative to the flux produced in a uniform background. 
We correct this result for halo substructure, and integrate over a 
large volume to determine the cosmological background from WIMP annihilation 
as a function of redshift.
The outline of this paper is as follows. In section 2, we define
a dimensionless flux multiplier $f$ that accounts for the enhanced
rate of two-body interactions produced by inhomogeneities in
the dark matter distribution, and determine its value for simple
virialised halos. In section 3 we calculate $f$ for cosmological volumes,
using analytic estimates of the halo mass function and halo concentrations,
and determine its redshift dependence.
Finally, in section 4 we study the contribution to $f$ from substructure
within virialised halos, and calculate $f$ for a set of realistic halos 
generated using a semi-analytic model of halo substructure. Throughout
this paper we assume a Lambda-CDM (LCDM) cosmology with a cosmological 
constant $\Lambda_0 = 0.7$, a matter density $\Omega_{{\rm m},0} = 0.3$ 
and a Hubble parameter $H_0 = h \times\ 100 {\rm km\,s^{-1}}$, 
with $h = 0.65$.

\section{Annihilation Rates in Simple Halos}\label{sec:simple}

In current hierarchical models, cold dark matter is expected to
form centrally concentrated halos with a characteristic density profile.
Dense substructure is abundant within these halos, as a relic from earlier 
stages of the hierarchical merging process. Since the annihilation flux is 
quadratic in the density, these inhomogeneities will increase the flux from 
a halo of a given mean density. We begin by computing a dimensionless quantity 
that describes this enhancement. In the next section we 
will then study the evolution of this quantity with epoch.

\subsection{The Dimensionless Flux Multiplier}\label{ssec:fdef}

If dark matter consists of neutralinos with a mass $m_\chi$ and a 
velocity-averaged cross-section for annihilation 
${\langle \sigma v \rangle}$, then the annihilation flux produced 
within a volume $V$ will be 
\begin{equation}
\Phi \propto \frac{\langle \sigma v \rangle}{m_\chi^2} \int_V \rho^2 dV\,,
\end{equation}\label{fluxm0}
where $\rho$ is the local density of CDM. For non-relativistic particles,
${\langle \sigma v \rangle}$ is approximately independent of $v$
so the flux will just be proportional to $\rho^2$.

Since the rate depends quadratically on the density, the total rate 
from a given mass within a given volume will be higher if the dark matter is 
distributed inhomogeneously. We can study this enhancement by defining
the dimensionless flux multiplier
\begin{equation}
f(V) \equiv {{1}\over{{\bar \rho}^2 V}}\int_V \rho^2 dV\,,
\label{fluxm1}
\end{equation}
for a distribution within a volume $V$, where ${\bar \rho}$ is the average
density within this volume. This function can also be written as a 
mass-weighted density average:
\begin{equation}
f(V) = \int_V {{\rho}\over{\bar{\rho}}} {{dm}\over{M}} \,,
\label{fluxm2}
\end{equation}
where $M$ is the total mass within $V$. Clearly $f = 1$ for a 
homogeneous distribution, while for a power-law density profile
$\rho \propto r^{-\alpha}$, 
\begin{equation}
f = {{(3 - \alpha)^2}\over{3(3 - 2\alpha)}}
\label{powlaw}
\end{equation}
provided $\alpha < 1.5$. For $\alpha = 1.5$, 
integrating equation (\ref{fluxm1})
from $r_{\rm min}$ to $r_{\rm max}$ gives:
\begin{equation}
f = {3 \over 4}\,\ln(r_{\rm max}/r_{\rm min})\, ,
\end{equation}
so $f$ diverges logarithmically as $r_{\rm min}$ goes to zero.

\subsection{Analytic Profiles}

Two analytic density profiles are commonly used to fit
the spherically averaged properties of dark matter halos, 
the NFW profile (Navarro, Frenk \& White 1996, 1997),
and the Moore profile (Moore et al.\ 1998).  
We can specify these generically as 
\begin{equation}
\rho(r) = \rho_s\,r_s^{\alpha + \beta\gamma} / r^{\alpha}\,(r^{\beta} + r_s^{\beta})^{\gamma}\, ,
\end{equation}
with $\alpha = \beta = 1, \gamma = 2$  for the NFW profile and 
$\alpha = \beta = 1.5, \gamma = 1$
for the Moore profile. The total mass within radius $r$ is 
\begin{equation}
M(<r) = 4 \pi r_s^3 \rho_s m(r/r_s)\, , 
\end{equation}
with 
\begin{equation}
m(x) = \ln(1 + x) - x/(1+x)
\end{equation}
for the NFW profile and 
\begin{equation}
m(x) = 2/3\, \ln (1 +x^{1.5})
\end{equation}
for the Moore profile, while the mean density within this radius is simply 
\begin{equation}
{\bar \rho(x)} = 3 \rho_s m(x)/x^3 , 
\end{equation}
where $x \equiv r/r_s$\,.

In a cosmological setting, halos are virialised out
to a radius $r_v$ corresponding to an overdensity $\Delta_c$ 
of roughly 200 relative 
to the background. The concentration $c \equiv r_v/r_s$ of a halo describes 
the size of this radius relative to $r_s$\,. \footnote{Note that the 
concentrations derived by fitting the outer regions of simulated halos 
differ according to the profile used, with 
$c_{\rm NFW} \simeq 1.73\, c_{\rm Moore}$.} 
Calculating the flux multiplier over the virialised region of a halo, 
we get a function which depends only on $c$:
\begin{eqnarray}
f(c) &= &{{1}\over{{\bar \rho}^2 V}}\int_0^{r_v} \rho(r)^2 4 \pi r^2 dr\\
 &= &{{c^3}\over{3 m^2(c)}}\int_0^{c} {{1}\over{x^{2\alpha}\,(1 + x^{\beta})^{2\gamma}}}\, x^2 dx .
\end{eqnarray}
Thus for the NFW profile 
\begin{eqnarray}
f(c) &= &{{c^3}\over{3 m^2(c)}}\int_0^{c} {{dx}\over{(1 + x)^{4}}}\\
 &= &{{c^3 - c^3/(1 + c)^3}\over{9\,(\ln(1+c) - c/(1+c))^2}} ,
\label{eq:nfw_fc}
\end{eqnarray}
with a limiting value of $f(c) = 4/3$ for small $c$, 
whereas for the Moore profile,
\begin{equation}
f(c) = {{c^3}\over{3 m^2(c)}}\int_{x_{\rm min}}^{c} {{dx}\over{x (1 + x^{1.5})^{2}}}
\end{equation}
\begin{equation}
 = {{2 c^3}\over{9 m^2(c)}}\left[{1\over{1 + x^{1.5}}} + \ln\left({x^{1.5} \over {1 + x^{1.5}}}\right) \right]^c_{x_{\rm min}}. 
\label{eq:moore_fc}
\end{equation}
This expression diverges logarithmically as ${x_{\rm min}}$ goes to zero,
reflecting the fact that the flux from a pure $r^{-1.5}$ cusp is
infinite. In practice, annihilation would reduce the density
at the centre of the halo to a finite value even for pure CDM, while
in real halos baryons will also affect halo structure on small scales.
Furthermore, the simulations that indicate central cusps in halos can only
probe down to $r/r_v \simeq 0.01$, or $r/r_s \simeq 0.1$ 
(Power et al.\ 2002); below this
scale, it is possible that the slope of the profile continues to change 
(Taylor \& Navarro 2001). Thus in what follows we will consider Moore
profiles truncated at various inner cutoff radii, either 
$x_{\rm min} = 10^{-10}$ (profile M1 hereafter),  $x_{\rm min} = 10^{-2}$ 
(profile M2 hereafter), $x_{\rm min} = 10^{-1}$ (profile M3 hereafter),
or $x_{\rm min} = 1/3$ (profile M4 hereafter), with cores of constant density 
within the cutoff radius. (Adding a core modifies equations 
(\ref{eq:nfw_fc}) and (\ref{eq:moore_fc}) slightly, but
the difference is negligible for $x \gg x_{\rm min}$.)
The first cutoff corresponds the scale on which annihilation alone might 
truncate a pure power-law density profile 
(C\'{a}lc\'{a}neo-Roldan \& Moore 2000) and 
the third cutoff corresponds to the largest core consistent with 
the results of current high-resolution simulations, while the fourth
cutoff corresponds to the limit of the region within which baryons
might plausibly have flattened out the core in the halos of massive galaxies.

\subsection{Non-analytic Profiles}

As mentioned above, the simulations that provide evidence for a universal
density profile for CDM halos can only reliably resolve two decades of
radius below the virial radius. 
While the logarithmic slope appears to converge 
to a value between $-1.5$ and $-1$ in the inner regions, it is clearly
not a simple power-law in the outer regions, where it flattens 
substantially over a decade in radius. Thus, there remains the 
possibility that the inner slope of CDM halos, were it resolved
to smaller radii, would continue to change in value. A shallower inner
slope would help to explain the rotation curves of dark-matter dominated
dwarf galaxies (e.g.\ Blais-Ouellette, Amram, \& Carignan 2001; de Blok \& Bosma 2002), as well as the mass distribution of the Milky Way
within the solar circle (Binney \& Evans 2001).

One possible form for such a profile was proposed by Taylor and Navarro 
(2001, TN hereafter). 
In the outer regions it resembles the NFW or Moore profiles, 
while in the inner regions (below about $\log(r/r_v) = -1.5$), its
logarithmic slope decreases slowly to $-0.75$. An analytic fit to the 
inner slope is: 
\begin{eqnarray}
{{d\,\ln\rho}\over{d\,\ln x}} 
= -\,{{0.75 + 2.625\, x^{1/2}}\over{1.0 + 0.5\, x^{1/2}}} ,
\end{eqnarray}
where $x = r/r_{\rm TN}$, with $r_{\rm TN} = 5/3\, r_{s,{\rm NFW}}$.
We will calculate $f(c)$, 
which is convergent and well-behaved for this profile, 
from a direct numerical integration, although it can be computed to reasonable 
accuracy using the fitting formula above. From equation (\ref{powlaw}), 
the limiting 
value for small $c$ is $f(c) = 9/8$. Note that in its outer regions, the 
equilibrium density profile derived by TN goes to zero at a finite radius. 
This assumes an equilibrium that will never be achieved in cosmological 
setting, however, where halos will continue to accrete material onto their 
outer parts. Thus, we assume that beyond $x_{\rm TN} = 1$, the profile drops
off as $r^{-3}$, as in the NFW and Moore fits.

\subsection{Comparison Between Profiles}

\begin{figure}
\centerline{\psfig{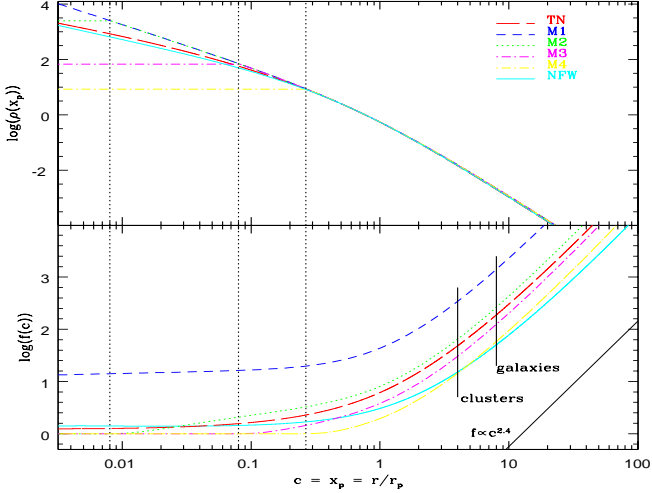}}
\caption{The dimensionless flux multiplier $f$ as a function
of concentration $c$, for the various density profiles shown in the top panel. 
The (solid) vertical lines
indicate typical concentrations for galaxies and
clusters at the present day, while the dashed lines indicate the
size of the constant-density core in profiles M2, M3 and M4. See text
for a description of the profiles. A line of slope 2.4 is shown in the bottom 
right-hand corner for comparison.
}\label{fig:1}
\end{figure}

Figure \ref{fig:1} shows the density profiles (top panel) and $f(x)$ 
(bottom panel) for the six profiles 
discussed above, including four versions of the Moore profile with 
different constant-density cores (within radii indicated by the horizontal 
dashed lines). In order to compare the different
profiles, they have been plotted in terms of $x_p \equiv r/r_p$, the
radius relative to the point $r_p$ at which the circular velocity peaks,
and normalised so that the density is the same at $r_p$. 

In the top panel, we see that all the density profiles aside from M4
are in excellent
agreement for $x_p \ge 0.1$, the region probed by current simulations.
Existing simulations are roughly compatible with any of the fits shown here.
The differences in the inferred central behaviour of halos are based on 
different interpretations of the density profile very close to the current 
numerical resolution limit. 

The bottom panel shows the effect of this uncertainty on $f(c)$, where $c$
is measured relative to $r_p$, as in the top panel. We see that 
in each case, the flux multiplier increases with concentration, reflecting 
the contribution from the dense central core. To first approximation, the
dependence on $c$ is independent of profile, while the inner slope of the 
halo profile determines the overall normalisation of $f(c)$. 
For the Moore profile it is 2--25 times larger than for an NFW profile, 
while for the shallower TN profile it is
comparable to a Moore profile with a core between 1\% and 10\% of $r_s$ 
in size.
The solid vertical lines on the plot indicate typical
values of $c$ for present-day galaxies and clusters.
For galaxy halos, $f(c)$ is typically 50 for an NFW profile, 200 for
the TN profile or profile M2, and 1200 for the most extreme Moore profile, M1. 
For all profiles it scales with concentration as roughly $c^{2.4}$ in this
range of $c$, 
as indicated by the solid line in the corner of the figure.
 
This figure indicates the importance of the inner region in determining
the total flux from annihilating dark matter. If recent simulations are 
correct, it seems likely that this flux may be 2--4 times larger than that 
predicted for a pure NFW profile; in the extreme case of a Moore profile 
with an inner cusp limited only by self-annihilation, 
the total flux is 25 times the NFW
value. Since the dependence on $r/r_p$ is similar for all profiles, in
what follows we will assume the NFW form for $f(c)$, and allow that
the flux may be a constant 2--25 times greater than this.


\section{Average over Cosmological Volumes}\label{sec:cosmo}

\subsection{Local Contribution}

To calculate the background flux from annihilations within a large
volume, we need to add the relative contributions from regions of different
density within this volume. Following the Press-Schechter approximation
(Press \& Schechter 1974), we can consider all the mass in a given
(physical) volume $V$ to be contained in virialised halos of some 
(possibly very small) mass. We make the further approximation that
these halos are spherical, with a fixed virial overdensity $\Delta_c$ 
relative to the critical density, and have universal density profiles with 
a concentration $c = r_v/r_s$ that depends only on their mass. 
If the volume contains $V\, (dn(M)/dM)\, \Delta M$ halos in the mass range
$M$ to $M + \Delta M$, then from equation (\ref{fluxm1}), 
the total flux multiplier for the volume will be
\begin{equation}
{f(V)} = {{1}\over{\rho^2\, V}}\, \int \left(V\, {{dn(M)}\over{dM}}\, dM\right)\, {\bar{\rho}}\, M\, f(c(M))\, .
\end{equation}
where $\rho$ is the mean density of dark matter within $V$,
${\bar{\rho}} = \Delta_c\,\rho_c$ is the average density of bound halos,
and $f(c(M))$ is the flux multiplier for halos of concentration $c = c(M)$.
Since $\rho = \Omega\,\rho_c$, we can rewrite the dimensionless 
flux multiplier for large volumes as:
\begin{eqnarray}
f(V) &=& \int f(c(M))\, {{\Delta_c\, \rho_c}\over{\Omega\, \rho_c}}\, {{M}\over{\rho}}\, {{dn(M)}\over{dM}}\, dM \\
 &=& {{\Delta_c}\over{\Omega}}\, \int f(c(M))\, {{dF(M)}\over{dM}}\, dM
\label{locvol}
\end{eqnarray}
where $F(M)$ is the fraction of the universe in virialised halos of mass
$M$ or larger. 

$F$ can be estimated using the Press-Schechter formalism:
\begin{eqnarray}
 {{dF_{\rm PS}(M)}\over{dM}}\, dM &=& {{dF_{\rm PS}(\nu)}\over{d\nu}}\, d\nu,\\
 &=& \left({2\over{\pi}}\right)^{1/2}\, \exp\left({{-\nu^2}\over{2}}\right)d \nu
\end{eqnarray}
with $\nu \equiv \delta_c/\sigma(M)$, where $\delta_c$ is the critical overdensity
and $\sigma(M)$ describes the power spectrum of density fluctuations. 
Recent simulations 
(e.g.\ Jenkins et al.\ 1998) have suggested, 
however, that
this mass function may overestimate the number of halos near $M_*$, the
characteristic mass for which $\sigma(M) = D(z)\,\delta_c$ (where $D(z)$ is the 
linear growth factor at redshift $z$). An alternative
mass function, proposed by Sheth and Tormen (1999) and based on the ellipsoidal 
collapse model, is:
\begin{equation}
 {{dF_{\rm ST}(\nu)}\over{d\nu}}\, d\nu = 2\,A\,\left(1 + {1\over{\nu^{2q}}}\right){{dF_{\rm PS}(\nu)}\over{d\nu}}\, d\nu,
\end{equation}
with $A = 0.3222$ and $q = 0.3$. 

Since both these mass functions
have a power-law behaviour at low masses, where concentrations and thus
$f(c)$ are larger, it is not clear the integral in equation (\ref{locvol}) 
will converge as we
include the contribution from smaller and smaller halos. (Although the
total mass per unit volume must converge, $f$ within a given volume 
need not converge, as in the case of a pure $r^{-1.5}$ density profile.)
In practice, we expect baryonic phenomena to complicate structure
formation on small mass scales, and annihilation itself will also limit
the contribution from very dense material.
Thus, we will truncate the integral at some limiting mass,
and consider the behaviour of $f$ as a function of this mass limit.

The concentration of a halo should reflect the density of the universe at
the time when it assembled the material now in its central core. 
There are several predictions of the concentration-mass relation
(Navarro, Frenk \& White 1997; Bullock et al.\ 2001; 
Eke, Navarro, \& Steinmetz 2001; Wechsler et al.\ 2002), 
based on this interpretation. We will
use the most recent analytic model, 
that of Eke, Navarro, \& Steinmetz (2001 -- ENS hereafter), to calculate the 
dimensionless flux multiplier, though we expect similar results from 
the other models. We use the public code supplied by the authors to 
calculate concentrations, and integrate the expressions above numerically.
Note that one point of uncertainty in all these models
is what minimum concentration to assign to halos that have just formed. 
The ENS code, for instance, predicts concentrations of 1 or less for
very massive or very high-redshift halos, but the least concentrated 
halos actually found in their
simulations have $c \simeq$ 2--3.
For recently formed objects the universal profile may not be well established,
so any uncertainty produced by varying the minimum concentration reflects
the limitations of the analytic model.

The top panel of figure 2 
shows the dimensionless flux multiplier contributed by halos 
over some limiting mass, at various epochs in a LCDM cosmology, plotted
as a function of the mass limit. The solid lines show the results for
the PS mass function, while the dashed lines show the results for the
ST mass function. We see that in either case, the total flux is 
more or less convergent as we include smaller and smaller masses;
at $z = 0$, it only varies by a factor of 2, for instance, for a 
limiting mass anywhere between $10^{10}\,M_\odot$ and $10^{3}\,M_\odot$. 
The value is also 
reasonably independent of the mass function used; it is at most 2 times
larger for the PS mass function. The uncertainties in halo concentrations
are only important at high redshift, when many halos have recently formed.
We have actually plotted two sets of curves in figure 2, one with a minimum
concentration of 1 (the lower set of lines at each redshift) and one with 
a minimum concentration of 2 (the upper set), but the difference between
them is negligible for $z < 10$. Finally, the total flux multiplier 
increases with time, reflecting the progressive growth of structure.
Of course the density of the universe is decreasing at the same time,
so the flux relative to that at the present day is larger at high redshift;
this is shown in the bottom panel, where we have used the present-day density
to normalise $f$. This increased flux will partly compensate for
the decreased volume element at high redshift, as discussed below.

\begin{figure}
\centerline{\psfig{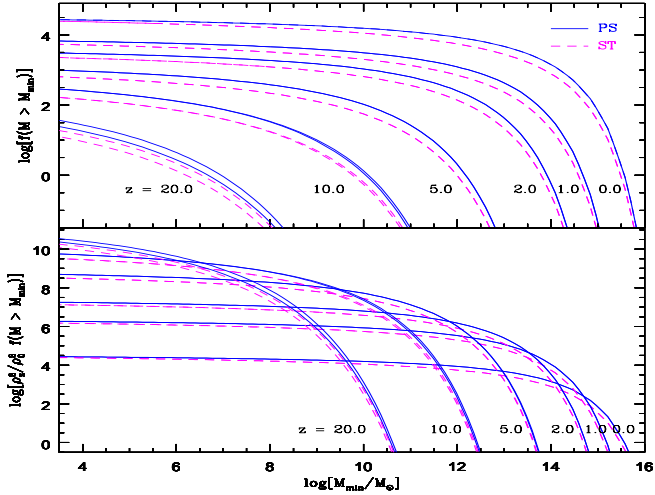}}
\caption{(Top panel) The dimensionless flux multiplier for large 
volumes $f(V)$ 
as a function of the minimum halo mass considered, at different
redshifts. The line styles indicate the mass functions assumed,
while the upper and lower lines visible at high redshift indicate
the effect of uncertainty in the halo concentrations for recently formed objects.
(Bottom panel) The flux multiplier relative to the present-day density
of the universe. NFW halo profiles and a LCDM cosmology are assumed.
}\label{fig:2}
\end{figure}

\subsection{The Universal Flux Multiplier}\label{ssec:univ}

In previous section we have derived the flux multiplier for a 
local volume, large enough to contain a representative sample of halos,
but small enough that evolutionary effects are negligible. In a similar 
fashion we can extend the definition of $f$ to entire observable universe. 
Consider the expression for $f$ integrated over volume elements $dV(a)$ 
at different scale factors $a$, to some maximum $a_{\rm m}$
\begin{eqnarray}
f_{\rm univ}(a_{\rm m})\hspace{-2mm}&\equiv&\hspace{-2mm}{{1}\over{V(a_{\rm m})\,{\bar{\rho}(0)^2}}}\int_1^{a_{\rm m}} \rho(a)^2 dV(a)\left| {{dt}\over{da}}\right|\\
\hspace{-2mm}&=&\hspace{-2mm}{{1}\over{V(a_{\rm m})\,{\bar{\rho}(0)^2}}}\int_1^{a_{\rm m}}
{\bar{\rho}}(a)^2f(a)dV(a)\left| {{dt}\over{da}}\right|,
\end{eqnarray}
where we have chosen to normalise by the present-day density of the universe,
${\bar{\rho}}(0)$, $f(a)$ is given by equation (\ref{locvol}), 
and the last term corrects for the time dilation at a given epoch.
Since $a = (1 + z)^{-1}$, this is equivalent to the following integral over 
redshift:
\begin{eqnarray}
f_{\rm univ}(z_{\rm m}) &=& {{3\,(1 + z_{\rm m})^3}\over{{\bar{\rho}}(0)^2}}\int_0^{z_{\rm m}}{{{\bar{\rho}}(z)^2f(z)\,dz}\over{(1 + z)^4}}\left| {{dt}\over{dz}}\right|\,,\\
{\rm with}\hspace{10mm}\nonumber\\
\left| {{dt}\over{dz}}\right| &=& (1 + z)^{-1}\,(\Omega_{{\rm m},0}\,(1 + z)^3 + \Lambda_0)^{-1/2}\,, 
\end{eqnarray}
for the LCDM cosmology considered here. 

\begin{figure}
\centerline{\psfig{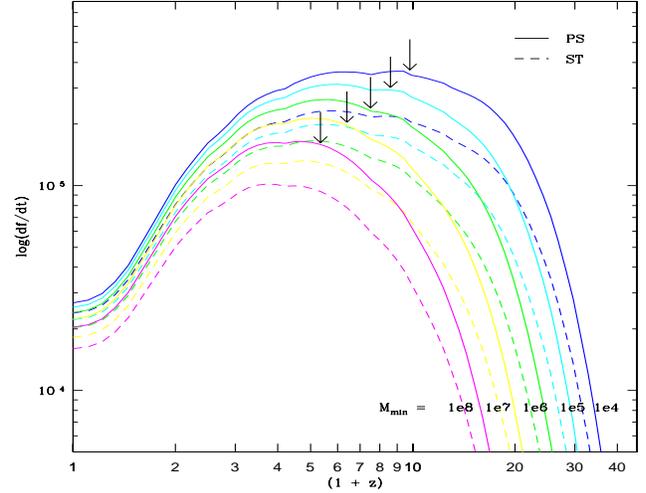}}
\caption{The relative contribution to the flux multiplier, 
per interval of time. The four sets of
lines indicate the contribution from mass functions extending down to the 
minimum masses indicated. The line styles indicate the results
for PS and ST mass functions. A LCDM cosmology is assumed. 
Arrows indicate the redshift at which
$M_* = M_{\rm min}$, i.e.\ $\sigma(M_{\rm min}) = D(z)\,\delta_c$.
}\label{fig:3}
\end{figure}

One quantity of interest is the relative contribution to the flux 
per interval of time, 
$df_{\rm univ}/dt = (df_{\rm univ}/dz) \left| {dz/dt}\right| $, 
which is plotted as a function of redshift in figure \ref{fig:3}. 
The four sets of
lines indicate the contribution from mass functions truncated at the
different minimum masses indicated. The line styles indicate the results
for PS and ST mass functions. We see that the relative contribution
peaks at quite early redshifts, roughly the epoch when the smallest halos
considered have formed, as indicated by the arrows. 
At higher redshift, it drops off sharply, because
there are very few halos over the mass limit considered, while at lower
redshift it drops off more gradually, as small, dense systems 
merge to form larger systems of lower density. The latter drop-off may
be overestimated, however, since the cores of dense halos may not be 
disrupted by merging, but may survive instead as distinct substructure within 
larger systems. We consider this effect in the next section. 

\section{Annihilation Rates in Realistic Halos}

In cosmologies dominated by CDM, large halos assemble hierarchically
through successive mergers between smaller objects. This merging
process is inefficient, and leaves many undigested remnants, or subhalos,
orbiting within a galaxy or cluster halo. The properties of these
objects have been studied extensively in several recent, high-resolution
SCDM and LCDM simulations (Ghigna et al.\ 1998; Klypin et al.\ 1999; 
Moore et al.\ 1999; Springel et al.\ 2001), and can also be understood using 
analytic and semi-analytic models (Bullock, Kravtsov, \& Weinberg 2000; 
Taylor 2001; Benson et al.\ 2002; Somerville 2002). 
The latter have the advantage
that the properties of the underlying halos can be specified directly,
and are independent of resolution and numerical effects. In this section
we use a recently developed semi-analytic model of halo formation 
(Taylor \& Babul in preparation; see also Taylor 2001) to determine the
flux multiplier for a realistic halo with substructure, at various different 
epochs.

\subsection{The Contribution from Halo Substructure}

From equation \ref{fluxm2}, 
the relative contribution to the flux multiplier from an individual 
subhalo within a larger halo is proportional to 
$\rho_{\rm sh} M_{\rm sh} f'(c_{\rm sh}) / \rho_{\rm bg} M_{\rm bg} f(c_{\rm bg})$, where
the subscript $sh$ indicates the subhalo and the subscript $bg$ indicates the 
average for the background system in which it resides. This implies 
that if subhalos are substantially denser or more concentrated than the
main system, they can dominate $f$, even though they
only contribute to a small fraction of the total mass of the system.
This is particularly true since tidal stripping will remove the
lowest-density material from substructure, increasing the relative
contribution from the material which remains bound.

More precisely, if we decompose the density of CDM within a halo into 
two components,
a smooth background following the basic profiles given in section 2,
and a set of subhalos which have been stripped to varying degrees, then
\begin{equation}
\rho({\mathbf x}) = \rho_{\rm bg}({\mathbf x}) + \sum_i \rho_{{\rm sh},i}({\mathbf x}).
\end{equation}
The contribution to $f$ then consists of several terms
\begin{eqnarray}
f\hspace{-2mm}&=&\hspace{-2mm}{{1}\over{{\bar \rho_{\rm bg}}^2 V_{\rm bg}}}\int_{V_{\rm bg}} (\rho_{\rm bg} + \sum_i \rho_{{\rm sh},i})^2\,dV\\
\hspace{-2mm}&=&\hspace{-2mm}{{V_{\rm bg}}\over{M_{\rm bg}^2}}\int_{V_{\rm bg}} (\rho^2_{\rm bg} + \rho_{\rm bg} \sum_i \rho_{{\rm sh},i} + \sum_i \rho^2_{{\rm sh},i})\,dV\\
\hspace{-2mm}&\simeq &\hspace{-2mm}f(c_{\rm bg}) + {{V_{\rm bg}}\over{M_{\rm bg}^2}}\left( \sum_i M_i \rho_{\rm bg}({\mathbf x_i}) 
+ \sum_i {{M_i^2}\over{V_i}} f'(c_i)\right),
\label{subst3}
\end{eqnarray}
where $M_{\rm bg}$, $c_{\rm bg}$ and $V_{\rm bg}$ are the mass, concentration and volume of the 
main system, and $M_i$, $c_i$ and $V_i$ are the corresponding properties of 
each subhalo. In deriving the final result, we have assumed that the
subhalos do not overlap, and that each one is small compared to the main 
system, so that $\rho_{\rm bg}$ is more or less constant within the volume 
covered by a given subhalo. The function $f'(c)$ is the dimensionless flux 
multiplier defined in section 1, modified to account for the effects of tidal 
heating and stripping, which may change the density profile of the subhalos 
as they orbit within the main system. We derive an expression for this
function in terms of the original profile and the amount of mass lost below.

\subsection{The Flux Multiplier for Subhalos}

The density profiles considered in section (\ref{sec:simple}) 
characterise isolated halos; 
interactions between halos or mergers with larger systems will modify 
these profiles through a combination of tidal heating and tidal 
mass loss. Numerical simulations show that tidal effects tend to
strip mass off the outer regions of a halo preferentially, preserving
the central density of the halo until it has lost a large fraction of its
original mass. Hayashi et al.\ (2002) propose the following modification
of the density profile to describe this process:
\begin{equation}
\rho'(r) = {{f_t}\over{1 + (r/r_t)^3}}\, \rho(r)\, ,
\label{hayashi1}
\end{equation}
where $r_t$ is the tidal radius to which the halo is stripped, and 
$f_t$ is the reduction in the central density. For a halo with an NFW
profile and an original concentration of 10, they find that these 
parameters are related to the fraction $m_{\rm b}$ of the original mass 
still bound to the halo by
\begin{eqnarray}
\log (r_t) &\simeq& 1.02 + 1.38 \log (m_{\rm b}) + 0.37 (\log (m_{\rm b}))^2\\
{\rm and} \nonumber\\
\label{hayashi2}
\log (f_t) &\simeq& -0.007 + 0.35 \log (m_{\rm b}) + 0.39 (\log (m_{\rm b}))^2\nonumber\\
&&+ 0.23 (\log (m_{\rm b}))^3 .
\label{hayashi3}
\end{eqnarray}

Using this modified density profile, we can calculate $f'$, the flux multiplier
for tidally stripped halos. This should be a function of $c_0$, the original
concentration of the halo, and $c_t \equiv r_t/r_s$, the concentration
of the stripped system. In practice, we shall consider $f'(c_t; c_0 = 10)$
since this was the profile modeled by Hayashi et al.\ (2002).
In defining $f'$, we also need to choose an appropriate volume over which
to integrate. Since the modified density profile in equation (\ref{hayashi1}) 
is not 
truncated completely at $r_t$, we could continue to integrate over the 
original volume; in this case, $f'(c)$ would rise sharply (roughly as 
$(r_t/r_{t,0})^{3}$) as the mass of the halo is confined to a smaller and
smaller region of the original volume. Since $r_t$ roughly defines 
the region in which the subhalo density profile dominates over the 
background, the approximation made in deriving equation (\ref{subst3}) 
will be more accurate if we define $f'$ over the volume within 
$r_t$. In any case, in this section we will show the results of the 
calculation for several possible choices of volume, to clarify the
effect on $f'$. 

\begin{figure}
\centerline{\psfig{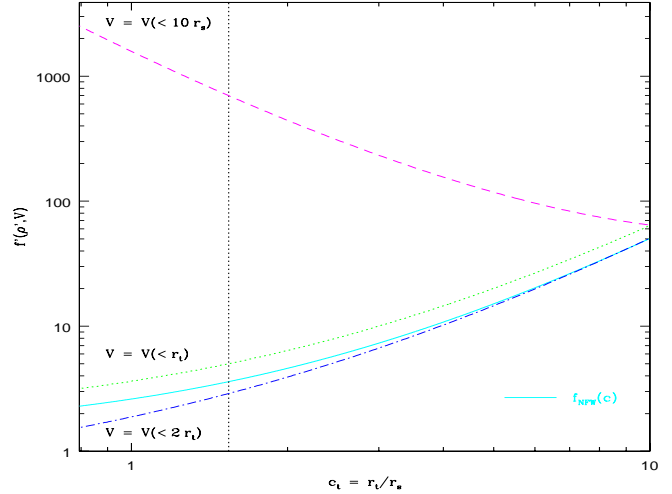}}
\caption{The dimensionless flux multiplier $f'$ 
calculated over various volumes, for stripped systems with 
an original NFW profile of concentration $c_0 = 10$, 
as a function of their stripped concentration $c_t \equiv r_t/r_s$. 
The dashed lines show $f'$ calculated within the original tidal
radius $r_{t,0} = 10\,r_s$, the dotted line shows $f'$ for the volume within
$r_t$, and the dot-dashed line shows $f'$ for the volume within
$2\,r_t$. The solid line shows $f$ for the original density profile,
truncated at different radii.}\label{fig:4}
\end{figure}

Figure \ref{fig:4} shows $f'(c_t; c_0 = 10)$ for the volume within 
$r/r_s = c_0 = 10\,r_s$ (dashed line), $r = r_t$ (dotted line), and 
$r = 2\,r_t$ (dot-dashed line), with $f(c_t)$ shown for comparison (solid line). 
The truncated functions differ from the original one at $c_t = c_0 = 10$
because the density profile given by equation (\ref{hayashi1}) is different 
from the 
unstripped one even when $r_t/r_s = c_0$. We see however that if we consider 
the volume within $r_t$ (dotted curve), $f'(c_t; c_0 = 10) \simeq 1.3\,f(c_t)$ 
to good approximation. We will assume that this result remains true for 
halos with larger or smaller original concentrations, 
provided $c_t < 0.9\, c_0$, and use this to calculate the 
contribution to the total flux from evolved subhalos.

\subsection{Results for LCDM Halos}

Using equations (\ref{subst3}) -- (\ref{hayashi3}), we can calculate 
the total flux multiplier
for composite systems with substructure. To do so, we use a
semi-analytic model of halo formation (Taylor \& Babul 
in preparation; see also Taylor 2001) 
to determine the number, masses, density profiles
and positions of subhalos within a set of hierarchically formed halos. 
We have followed halo substructure down to 
$\simeq 10^{-5}$ times the mass of the main system. Since the spectrum 
of substructure is fairly steep at low masses, the total flux multiplier 
may depend on this cutoff, so we will consider our results 
as a function mass limit, as in section (\ref{sec:cosmo}), 
and extrapolate downwards to estimate the total flux multiplier for smaller
limiting masses.

\begin{figure}
\centerline{\psfig{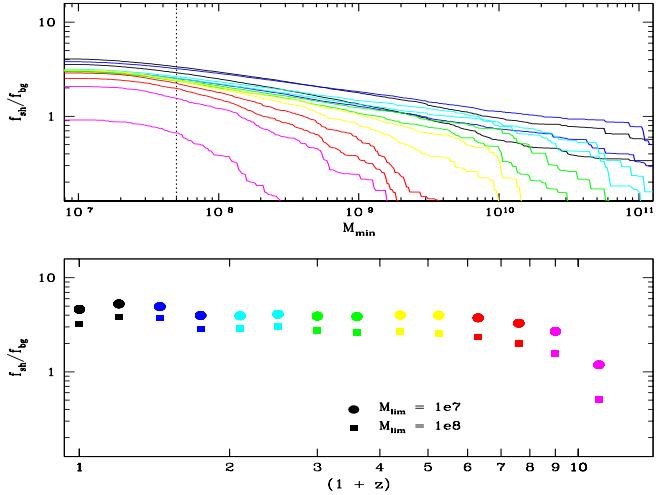}}
\caption{Top panel: the contribution to the flux multiplier contributed 
by substructure over a minimum mass $M_{\rm min}$, as a function of 
$M_{\rm min}$, relative to the flux multiplier $f_{\rm bg}$ expected for 
a smooth NFW halo without substructure.The solid curves are average results 
for a set of semi-analytic galaxy halos with a total mass of 
$1.6\times 10^{12}\,M_{\odot}$, in a LCDM cosmology. The dashed 
vertical line indicates the lower mass limit of the merger trees used. 
Bottom panel : as above, but showing the redshift dependence of the 
relative contribution for two specific mass limits, 
$M_{\rm lim} = 10^7\,M_{\odot}$ 
and $M_{\rm lim} = 10^8\,M_{\odot}$.}\label{fig:5}
\end{figure}

The top panel of figure \ref{fig:5} shows the relative contribution 
to the flux multiplier contributed by substructure over a minimum mass 
limit $M_{\rm min}$, as a function of $M_{\rm min}$, in models using 
NFW profiles. The contribution has been normalised to the 
value $f_{\rm bg}$ expected for a smooth NFW halo without substructure. 
The solid curves are average results for a set of 
semi-analytic galaxy halos with a total mass of $1.6\times 10^{12}\,M_{\odot}$,
in a LCDM cosmology. The dashed vertical line indicates the resolution limit
of the merger trees used; below this, the spectrum of subhalos becomes 
increasingly incomplete, particularly below $2\times 10^{7}\,M_{\odot}$.
At high redshift, where the halos are forming through major mergers around
this mass scale, the curves are steep so the flux estimate will depend 
sensitively on the mass limit. At low redshift, the effect of the 
mass limit is less important,
as most of the flux comes from more massive subhalos. In either case,
however, the total flux appears to have a power-law dependence on 
$M_{\rm lim}$ at low masses, so it is straightforward to extrapolate 
our estimates to lower mass limits.

Overall, we see that substructure makes an important contribution 
to total flux right up to the present day, when it exceeds the 
contribution from the background halo. 
The bottom panel shows the redshift dependence of the contribution 
more clearly,
for two different mass limits, $M_{\rm lim} = 10^7\,M_{\odot}$ 
and $M_{\rm lim} = 10^8\,M_{\odot}$. As expected, at high redshift 
the contribution from substructure increases somewhat faster as we lower 
$M_{\rm lim}$, reflecting the steep slope of the high-redshift
curves in the top panel, while at low redshift the increase is slower. 
If we extrapolate the
power-law dependence seen in the top panel to smaller values of $M_{\rm lim}$
such as $M_{\rm lim} \simeq 10^4$--$10^5$, we expect a net contribution from 
substructure of anywhere from 10 to 20 times the smooth background value 
for the NFW profile.

\begin{figure}
\centerline{\psfig{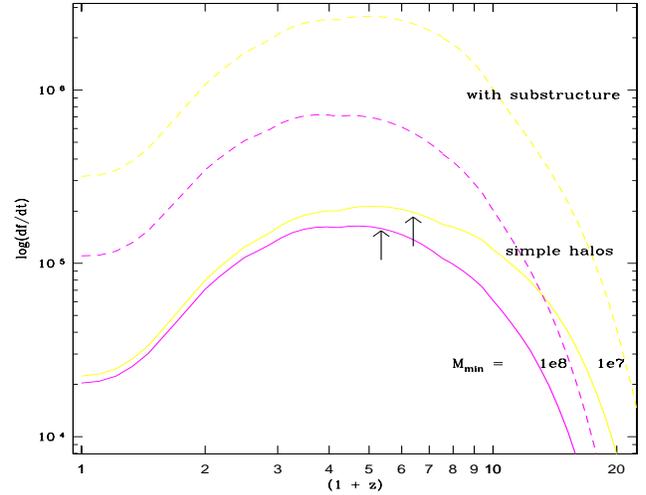}}
\caption{As figure \ref{fig:3}, but for two specific mass cutoffs 
(solid lines),
and including the effects of substructure on the total flux multiplier
(dashed lines). The arrows indicate the redshift at which
$M_* = M_{\rm min}$, as before.}\label{fig:6}
\end{figure}

We can determine the effect of substructure on the gamma-ray background
by correcting the relative contribution calculated in section 
(\ref{ssec:univ}) for the additional contribution from substructure
in shown in figure \ref{fig:5}; this is indicated by the dashed lines in 
figure \ref{fig:6}, for halos with NFW profiles, and mass 
cutoffs of $M_{\rm lim} = 10^7\,M_{\odot}$
and $M_{\rm lim} = 10^8\,M_{\odot}$. We see that when substructure is
taken into account, the flux multiplier still peaks just after the epoch 
when $M_* = M_{\rm lim}$ and declines at lower redshift, but both
the peak value and the value at $z = 0$ are substantially enhanced. 
We note that the peak and subsequent decline may partly compensate 
for the cosmological factor $\left| dt/dz \right|$
and for intervening absorption, which reduce the high-redshift contribution
to the gamma-ray background. We will investigate the detailed line
and continuum spectra predicted by this model in a subsequent paper.

\section{Discussion}

We have calculated the amount by which structure in dark matter on
subhalo, halo and cosmological scales amplifies the expected signal
from neutralino annihilation. The overall magnitude of the effect 
remains uncertain at several levels, some more important than others. 
Uncertainty in
the fundamental density profile which characterises halos, or more
specifically in the slope of their density profiles at small radii,
can change the flux multiplier for simple halos by a factor of 20--25, assuming
concentrations typical of galaxy halos (see figure \ref{fig:1}).
There is evidence in recent simulations for some excess mass in 
the inner regions, relative to the NFW profile (Power et al.\ 2002), 
so a reasonable estimate for the flux multiplier may be 3--4 times the NFW
value used for most of the calculations in this paper, although studies of 
the Milky Way (e.g.\ Binney \& Evans 2001) suggest less dark matter in its 
central regions than these profiles would imply. There is
no direct evidence for a pure Moore profile extending to very small
scales, although if this is the case the resulting flux from simple halos
will be 20--25 times the NFW value used here.

The mass function of dark matter halos, averaged over large volumes,
is also uncertain, particularly at the low mass end. The mass function 
determined from simulations is better fit by the form proposed by
Sheth and Tormen (ST) than by the traditional Press and Schechter (PS) 
mass function. Using the ST mass function to predict the background flux
reduces its amplitude by about 1.5 over PS. We assume
that the scale invariance of dark matter substructure will be broken
at very small masses, perhaps close to Jeans mass at recombination, for
instance. At a minimum, there should be some maximum density for
dark matter halos, or equivalently some maximum for $\sigma(M)$ at early
times. Varying the lower mass limits over a reasonable range 
($10^4$--$10^8\,M_{\odot}$) changes the flux by a factor of 2 or
so for simple halos, but the contribution from substructure may
increase this difference to as much as a factor of 40. 

Overall, these uncertainties combine to produce a total uncertainty 
of more than an order of magnitude in the flux multiplier.
A conservative model, at the bottom of this range, consists of
an NFW profile, with a ST mass function and a mass limit
of $10^6\,M_{\odot}$. For this model,
the present-day flux multiplier is $6\times 10^5$. We estimate that
a more likely model has a profile with more mass in its inner regions,
and lower limit to the mass function of $10^5\,M_{\odot}$. For this model,
the present-day flux multiplier is $5\times 10^6$. For an extreme model,
with a Moore profile limited only by annihilation, and a mass
limit close to the Jeans mass at recombination, the present-day flux 
multiplier will be $\simeq 1\times 10^8$.

It is worth noting that with these large flux multipliers, some 
neutralino candidates can already be ruled out. The 86 GeV
neutralino considered in Bergstr\"{o}m, Edsj\"{o}, \& Ullio (2001), 
for instance, 
produced almost 10\% of the gamma-ray background observed by EGRET 
at 1 GeV, assuming a flux multiplier of $\simeq 2\times 10^6$, while
a 166 GeV neutralino produced more than 50\% of the flux observed at 
10 GeV. Both these candidates are thus possible in our most conservative
model, while the higher-energy candidate is marginally excluded in our 
favoured model, and both are ruled out in our extreme model.

\section*{Acknowledgments}

The authors would like to thank E. Hayashi and his collaborators for 
making their results available prior to publication. We also thank 
A. Babul and J. Edsj\"{o} for useful discussions.
JET gratefully acknowledges funding from the Leverhulme Trust during
the course of this work.



\begin{thebibliography}{99}

\bibitem[Binney \& Evans(2001)]{2001MNRAS.327L..27B} Binney, J.~J., \& 
Evans, N.~W.\ 2001, MNRAS, 327, L27 

\bibitem[Blais-Ouellette, Amram, \& Carignan(2001)]{2001AJ....121.1952B} 
Blais-Ouellette, S., Amram, P., \& Carignan, C.\ 2001, AJ, 121, 1952 

\bibitem[Benson et al.(2002)]{2002MNRAS.333..156B} Benson, A.~J., Lacey, 
C.~G., Baugh, C.~M., Cole, S., \& Frenk, C.~S.\ 2002, MNRAS, 333, 156 

\bibitem{lbreview}
Bergstr\"om, L. 2000, Rept.\ Prog.\ Phys.\ 63, 793

\bibitem[Bergstr\"{o}m, Edsj\"{o}, \& Ullio (2001)]{2001PRL...87..25130} Bergstr\"{o}m, L., 
Edsj\"{o}, J., \& Ullio, P. 2001, Phys.\ Rev.\ Lett.\ 87, 251301

\bibitem[Bullock, Kravtsov, \& Weinberg(2000)]{2000ApJ...539..517B} 
Bullock, J.~S., Kravtsov, A.~V., \& Weinberg, D.~H.\ 2000, ApJ, 539, 517 

\bibitem[Bullock et al.(2001)]{2001MNRAS.321..559B} Bullock, J.~S., Kolatt, 
T.~S., Sigad, Y., Somerville, R.~S., Kravtsov, A.~V., Klypin, A.~A., 
Primack, J.~R., \& Dekel, A.\ 2001, MNRAS, 321, 559 

\bibitem[de Blok \& Bosma(2002)]{2002A&A...385..816D} de Blok, W.~J.~G.,\& 
Bosma, A.\ 2002, AAP, 385, 816 

\bibitem[C\'{a}lc\'{a}neo-Roldan \& Moore(2000)]{CM2000} C\'{a}lc\'{a}neo-Roldan, C., \& 
Moore, B.\ 2000, Phys.\ Rev.\ D62, 123005 

\bibitem[Eke, Navarro, \& Steinmetz(2001)]{2001ApJ...554..114E} Eke, V.~R., 
Navarro, J.~F., \& Steinmetz, M.\ 2001, ApJ, 554, 114 

\bibitem[Ghigna et al.(1998)]{1998MNRAS.300..146G} Ghigna, S., Moore, B., 
Governato, F., Lake, G., Quinn, T., \& Stadel, J.\ 1998, MNRAS, 300, 146 

\bibitem{Hayashi:2002}
Hayashi, E., Navarro, J.~F., Taylor, J.~E., Stadel, J., \& Quinn, T. 2002, 
ApJ, submitted (astro-ph/0203004)

\bibitem[Jenkins et al.(1998)]{1998ApJ...499...20J} Jenkins, A.~et al.\ 
1998, ApJ, 499, 20 

\bibitem[Klypin, Gottl{\" o}ber, Kravtsov, \& 
Khokhlov(1999)]{1999ApJ...516..530K} Klypin, A., Gottl{\" o}ber, S., 
Kravtsov, A.~V., \& Khokhlov, A.~M.\ 1999, ApJ, 516, 530 

\bibitem[Moore et al.(1999)]{metal99} Moore B., Ghigna S., Governato F., Lake G., Quinn
T., Stadel J., \& Tozzi P., 1999, ApJ, 524, L19

\bibitem[Moore et al.(1998)]{1998ApJ...499L...5M} Moore, B., Governato, F.,
Quinn, T., Stadel, J., \& Lake, G.\ 1998, ApJ, 499, L5

\bibitem{nfw96} Navarro J.F., Frenk C.S., \& White S.D.M., 1996, ApJ, 462, 563

\bibitem{nfw97} Navarro J.F., Frenk C.S., \& White S.D.M., 1997, ApJ,
490, 493

\bibitem{Power:2002}
Power, C., et al.\ 2002, MNRAS, submitted (astro-ph/0201544)

\bibitem[Press \& Schechter(1974)]{1974ApJ...187..425P} Press, W.~H., \& 
Schechter, P.\ 1974, ApJ, 187, 425 

\bibitem[Sheth \& Tormen(1999)]{1999MNRAS.308..119S} Sheth, R.~K.~\& 
Tormen, G.\ 1999, MNRAS, 308, 119 

\bibitem[Somerville(2002)]{2002ApJ...572L..23S} Somerville, R.~S.\ 2002, 
ApJ, 572, L23 

\bibitem[Springel, White, Tormen, \& Kauffmann(2001)]{2001MNRAS.328..726S} 
Springel, V., White, S.~D.~M., Tormen, G., \& Kauffmann, G.\ 2001, MNRAS, 
328, 726 

\bibitem[Taylor (2001)]{t2001} Taylor, J.~E.\ 2001, Ph.D. thesis, University of Victoira\ 2001


\bibitem[Taylor \& Navarro(2001)]{2001ApJ...563..483T} Taylor, J.~E., \& 
Navarro, J.~F.\ 2001, ApJ, 563, 483 

\bibitem[Ullio, Bergstr\"{o}m, Edsj\"{o}, \& Lacey(2002)]{ubbl} Ullio, P., 
Bergstr\"{o}m, L., Edsj\"{o}, J., \& Lacey, C. 2002, preprint 
(astro-ph/0207125)

\bibitem[Wechsler et al.(2002)]{2002ApJ...568...52W} Wechsler, R.~H., 
Bullock, J.~S., Primack, J.~R., Kravtsov, A.~V., \& Dekel, A.\ 2002, ApJ, 
568, 52 

\end{thebibliography}
\end{document}